\documentclass[conference]{IEEEtran}
\usepackage[]{graphicx}
\DeclareGraphicsExtensions{.eps}
 \graphicspath{{./Figures/}}
\usepackage[cmex10]{amsmath}
\usepackage[]{amsfonts}
\usepackage[tight,footnotesize]{subfigure}
\usepackage{balance}
\usepackage{xfrac}
\usepackage{enumerate}	
\usepackage{stfloats}

\begin{document}

\title{{Leveraging Deliberately Generated Interferences for Multi-sensor Wireless RF Power Transmission}}

\author{
\IEEEauthorblockN{
Raul G. Cid-Fuentes\IEEEauthorrefmark{1},
M. Yousof Naderi\IEEEauthorrefmark{2},
Rahman Doost-Mohammady\IEEEauthorrefmark{2}, \\
 Kaushik R. Chowdhury\IEEEauthorrefmark{2},
 Albert Cabellos-Aparicio\IEEEauthorrefmark{1} and~Eduard Alarc\'on\IEEEauthorrefmark{1},}

\vspace{0.2cm}
\IEEEauthorblockA{
\IEEEauthorrefmark{1}NaNoNetworking Center in Catalunya (N3Cat)
Universitat Polit\`{e}cnica de Catalunya, Spain\\
\IEEEauthorrefmark{2},Electrical and Computer Engineering Department
Northeastern University, USA \\
Email: rgomez@ac.upc.edu, naderi@coe.neu.edu, doost@ece.neu.edu,\\ krc@ece.neu.edu, acabello@ac.upc.edu, eduard.alarcon@upc.edu \vspace{-0.4cm}} \\

}

\maketitle

\begin{abstract}
Wireless RF power transmission promises battery-less, resilient, and perpetual wireless sensor networks. Through the action of controllable Energy Transmitters (ETs) that operate at-a-distance, the sensors can be re-charged by harvesting the radiated RF energy. However, both the charging rate and effective charging range of the ETs are limited, and thus multiple ETs are required to cover large areas. While this action increases the amount of wireless energy injected into the network, there are certain areas where the RF energy combines destructively. To address this problem, we propose a duty-cycled random-phase multiple access (DRAMA). Non-intuitively, our approach relies on deliberately generating random interferences, both destructive and constructive, at the destination nodes. We demonstrate that DRAMA optimizes the power conversion efficiency, and the total amount of energy harvested. 
Through real-testbed experiments, we prove that our proposed scheme provides significant advantages over the current state of the art in our considered scenario, as it requires up to 70\% less input RF power to recharge the energy buffer of the sensor in the same time. 
\end{abstract}

\section{Introduction}
Wireless RF power transmission is emerging as a promising approach to enable battery-less wireless sensor networks (WSNs)~\cite{zane-proceedings,naderi-RFMAC}. This technique aims to leverage RF energy harvesting~\cite{survey-sources}, which will allow controlled powering of nodes that may have insufficient residual energy in their batteries, or are unable to scavenge energy from the ambient environment at desired rates.
Given the relatively short charging range of one energy transmitter (ET), multiple ETs are required to cover large deployment areas~\cite{naderi-RFMAC} in WSN. The presence of multiple ETs reduces the average propagation distance to the energy harvesting sensors, and thus decreases the attenuation level of the energy waves and improves the RF power harvesting rates.  

\begin{figure}
  \centering
    \includegraphics[width=0.4\textwidth]{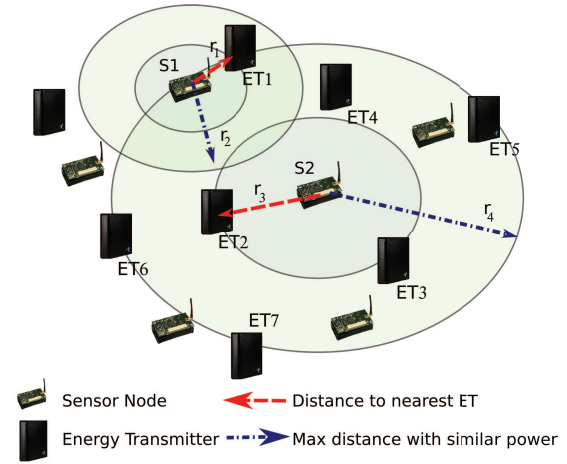}
 \caption{Example of a WSN powered by wireless RF energy transfer. The number of coinciding RF waves generated by ETs with similar magnitude depends on the distance from the sensor to the closest ET. Larger distances imply more coinciding sources.}
 \label{fig:interferers}
 \vspace{0.0 cm}
\end{figure}

Fig.~\ref{fig:interferers} shows a many-to-many power transmission in a WSN, where more than one ET delivers power to multiple sensors. In this network RF waves may interfere with each other when they are transmitted in the same medium. These interferences can be either constructive (i.e., the received power is larger than the average) or destructive (i.e., the received power is very low, or even zero) as shown in~\cite{naderi-RFMAC}, requiring multiple access techniques and medium access control (MAC) approaches for energy transfer among multiple ETs. 

The aim of existing MAC protocols for RF energy harvesting sensor networks with multiple ETs is to mitigate the impact of interferences. Along these lines, energy-on-demand (EoD) MAC protocol for in-band energy and data transfer in WSN by synchronizing the phase of the energy transmitters to maximize the power transfer at a given location was proposed in~\cite{naderi-RFMAC}. However, this phase maximization can be costly and complex to achieve in a many-to-many power transmission topology. Moreover, the power transmission is optimized for a single energy requesting node. Alternatively, the Powercast TX91501~\cite{powercast} implements FHSS and DSSS modulations, both of which aim to provide orthogonal channels among ETs, thus mitigating the effect of interference. 
Even though by using these modulations an arbitrary amount of ETs can be aggregated in the network, spread-spectrum approaches involve increasing the frequency spread of the energy signal (thereby, reducing the spectrum available for useful data) and require broadband energy harvesting hardware.

In this paper, we present a Duty-cycled RAndom-phase Multiple Access (DRAMA) scheme for wireless RF power transmission.  Our approach is based not only in handling the interferences of multiple ETs, but also to benefits from them to broaden the input power range of existing energy harvesters at the sensor nodes. For this, DRAMA relies on the fundamental assumption that efficiency is maximized when the input power varies in time as much as possible, since the energy harvesters operate with increasing efficiency as a function of the input power~\cite{P1110,kaushik-jetcas}.
 The key idea here is to generate time-synchronized duty-cycled bursts of energy (i.e., every ET transmits the burst at the same time), as these prove to be more efficient than a continuous transmission. Then,  we further increase the time-varying character of the received signal by selecting a random varying phase at each ET.

The main contributions of this paper can be summarized as:
\begin{itemize}
\item We introduce DRAMA and describe its operation.
\item We present an analytical model to characterize DRAMA.
\item We experimentally evaluate its performance using off-the-shelf circuits.
\end{itemize}

The rest of the paper is organized as follows. In Sec.~II, we overview the hardware and network considerations. In Sec. III, we present and analyze our DRAMA approach. In Sec. IV and Sec. V we numerically and experimentally evaluate the performance of DRAMA in a many-to-one environment.  Finally, we conclude our work in Sec. VI.

\section{Overview}
This section overviews the hardware and network architectures that are considered in this work.

\subsection{Network Model}

We consider a network composed of multiple ETs, which permanently transfer energy through RF waves to power a wireless sensor network, as shown in Fig.~\ref{fig:interferers}. Each node is equipped with an energy harvesting circuit, which converts RF energy into DC current. 
The received power at a sensor node located in an arbitrary location of the network is the result of the combination of the RF waves that are transmitted from multiple ETs. Due to the propagation path-loss, sensors will receive a large amount of power from their closest ETs, whereas the nodes located far from any ET will receive significantly less power that is generated by the small contributions of many ETs. 
We show an example of a network in Fig.~\ref{fig:interferers}. In the figure we observe that the sensors S1 and S2 are located at a distance $r_1$ and $r_3$ from their closest ETs, respectively. Then,  the distances $r_2$ and $r_4$ represent the maximum distance where an ET could generate an RF wave with similar power than the nearest ET. Notice that these distances depend on the distances $r_1$ and $r_3$. We see that due to the short distance $r_1$ between S1 and ET1, S1 only receives power from a single ET.
 Alternatively, S2 is located at a further distance from any ET, thus its maximum range is broadened. As such, it receives a power from ETs with indices ranging from 1 to 7.

\begin{figure}
  \centering
    \includegraphics[width=0.48\textwidth]{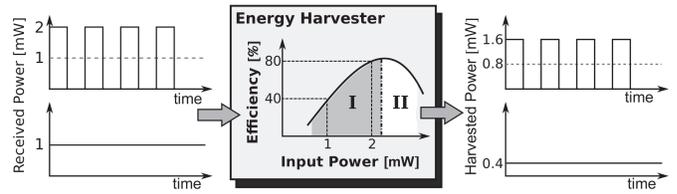}
 \caption{ Low and high power regions of operation of an energy harvester (denoted as I and II) and benefits of transmitting power with large peak-to-average power ratios in the region I. A duty-cycled transmission of energy is compared to a constant transmission.}
 \label{fig:example}
 \vspace{0.0 cm}
\end{figure}

\subsection{Hardware Considerations}
The non-linear behavior of semiconductor devices results in the dependency of the input impedance to the input power, such that the antenna and energy harvester impedances match only for a certain input power.
The impedance matching makes two distinguished regions in any real implementation~\cite{kaushik-jetcas}, as shown in Fig.~\ref{fig:example}:
Increasing efficiency for low input powers (denoted as region I) and decreasing efficiency for high input powers (denoted as region II).  
In region I, transmitting power in a time-varying manner leads to higher amounts of harvested energy~\cite{cttc-wpt}.
 On the contrary, in region II the power conversion efficiency at the high power range decreases with the input power~\cite{P1110}, and a low peak-to-average received power ratio improves the efficiency of the energy harvester.

As an example, Fig.~\ref{fig:example} also compares two power transmission schemes, namely continuous and duty-cycled (with a peak-to-average power ratio of 2) with equal average received power of 1~mW. Without loss of generality, we assume a $40\%$ efficiency for $1$~mW input power and a $80\%$ efficiency for $2$~mW input power.  Accordingly, in the case of continuous power transmission, a constant power of $0.4$~mW would be harvested, while in the case of duty-cycled power transmission the harvester provides a duty cycled instantaneous power of $0$~mW for an input power of 0~mW and $1.6$~mW for an input power of $2$~mW. Overall, this results in $0.8$~mW average harvested power. In contrast to this example, if the efficiency curve decreases with the input power, receiving a constant power provides a larger efficiency.

\section{DRAMA Scheme}

\begin{figure}
  \centering
    \includegraphics[width=0.48\textwidth]{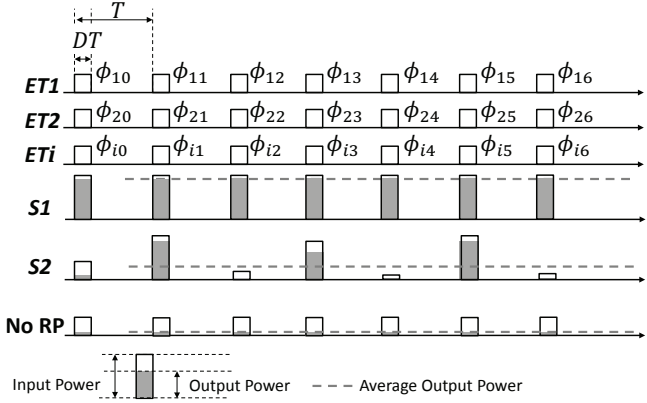}
 \caption{DRAMA scheme. Each ET propagates RF waves with random phases $\phi_{ik}$ in a duty-cycled manner with period $T$ and duty cycle $D$.}
 \label{fig:mac-scheme}
 \vspace{0.0 cm}
\end{figure}

Our proposed Duty-cycled RAndom phase Multiple Access (DRAMA) exploits the two energy regions in order to maximize the efficiency of the energy harvester at each node. In particular, it
 generates time-varying power at the input of the sensors that receive low power levels and to generate constant input power at the sensors that receive high power levels. For this, our scheme leverages multiple transmissions at the sensors that receive low power levels (i.e., sensors located at further distances from any ET receive low power and this power is generated by the combination of multiple ETs) and single transmission at the sensors that receive high power levels (i.e., sensors located nearby an ET receive the high power from their nearest ET, thus neglecting the combination from further ETs).
To generate time-varying input power, ETs transmit time-synchronized bursts of power in a duty-cycled manner (each {\em on} duration is here referred as a \emph{burst}). Each burst is a single frequency, continuous sine wave RF transmission, generated with a randomly selected phase $\phi_{jk}$. At the start of every \emph{on} time, the ETs select a random, different phase. Fig.~\ref{fig:mac-scheme} shows a time-diagram of the DRAMA scheme, where S1 and S2 are nodes in Fig.~\ref{fig:interferers}, whereas No RP stands for a generic multiple access method that ensures perfect channel orthogonality, such as FDMA or DSSS.

By employing this scheme we modulate the peak-to-average power ratio in two stages:
First, the duty-cycled transmission increases the peak-to-average power ratio of all input powers.
Second, the combination of multiple RF waves with random phases deliberately generates both destructive and constructive interferences. These interferences modulate the overall received power at the sensor locations, thus increasing the peak-to-average power ratio. 
Given that the time-varying power is generated by intentionally interfering the RF waves from the ETs, only sensors that receive power as the combination of multiple ETs receive a time-varying power. Accordingly, it is shown in Fig.~\ref{fig:mac-scheme} that S1 receives power only from one source (ET1), thus harvesting from a low peak-to-average power ratio signal. Alternatively, S2 receives power from many ETs (ET1 to ET7), thus harvesting from a very large peak-to-average power ratio signal.

The period, $T$, in DRAMA depends on the design and circuital  properties of the energy harvesting. These parameters must be experimentally adjusted for any given technology. In particular, we have experimentally found that for a P1100 Powercast harvester~\cite{P1110}, values of $T$ in the order of 10~ms to 100~ms provide a good performance.
This selected time range  is important as the burst must be significantly larger than the synchronization time, and the transient time of the energy harvesters at the receiving node.
Note that synchronization errors between ETs can be kept several orders of magnitude below the time-scale of DRAMA. In particular, a precision of a few tens of microseconds can be achieved using inexpensive off-the-shelf software over wireless~\cite{synch-survey}. This precision can be improved to 200~ns if ETs integrate typical GPS receivers~\cite{GPS-timing}.

\subsection{Analytical Model of DRAMA Performance}

\subsubsection{Combined Input Power}

We first model the input power at the sensor node destination. We denote $m$ as the number of ETs and $n$ as the number of sensors that are deployed over the networking area.
Without loss of generality, we locate the sensor node under study at the center location of the Cartesian plane. Then, we denote $P_{T}$ as the average power that is transmitted from an ET ($P_{Ti}$ if particularized by the $i$-th ET). 

We utilize the concept of signal $s(t)$ as the squared-root of the instantaneous input power, $p(t) = |s(t)|^2$, in W$^{1/2}$ units. Notice that the concept of the signal does not directly correspond to voltage, since the signal is $s(t) = \sqrt{v(t)i(t)}$, where $v (t) $ is the voltage drop across the terminals, and $i(t)$ is the current flowing through them.
According to operational method of the DRAMA scheme, \emph{the power is transmitted in bursts of energy and each ET transmits at the same time with same frequency and with random phase}, the duty cycled signal that is generated at a given ET can be described as the addition of independent bursts of RF signal delayed in time:
\begin{equation}
s(t) = \sum_{k=0}^\infty \sqrt{\frac{P_{T}g(r)}{D}} \cos \left( 2 \pi f_0 t +\phi_{k} \right) \Pi \left( \frac{t - kT}{DT} \right),
\end{equation}
where $k$ refers to the burst, $P_T$ is the average transmitted power, $g(r)$ stands for the path loss, which is a function of the distance $r$, $D$ is the duty cycle, $f_0$ stands for the carrier frequency, $\phi_{k} \in [0,2\pi)$ refers to the phase shift of the $k$-th burst, $\Pi (x)$ is the rectangular function (defined as $\Pi = 1$ if $0 \leq x \leq 1$ and $\Pi = 0$ otherwise), finally, $T$ refers to the time period of the duty cycle.
Moreover, the signal parameters $D$, $f_0$ and $T$ are fixed for all the ETs, such that each ET concurrently transmits in-time and in-frequency. On the contrary, the sequence of phase shifts, $\phi_{k}$, is random in-time and different ETs have different sequences.
 
To reduce further notation, we denote the received power from a given ET as $P_i = P_{Ti} g(r) $. 
Also, we focus on a specific burst time and use phasor notation, such that the signal that a sensor receives from a the $i$-th ET is simplified to:
\begin{equation}
s = \sqrt{P_i / D} e^{j\phi_i}.
\end{equation}

The received signal, composed as the combination of concurrent $m$ ETs at a given burst time, is a sine signal of same carrier frequency, amplitude $\sqrt{p_R}$ and phase $\theta$, which is given by:
\begin{equation}
s_R = \sum_{i=1}^m \sqrt{P_i / D} e^{j\phi_i} = \sqrt{p_R} e^{j \theta},
\end{equation}
where $p_R$ is the instantaneous received power and can be easily obtained as:
\begin{equation}
p_R = D^{-1} \sum_{i=1}^m P_i +  D^{-1} \sum_{\substack{1 \leq i,j \leq m \\ i \neq j}} \sqrt{P_i P_j}  cos(\phi_i - \phi_j).
\label{eq:p_sum}
\end{equation}
As observed,
the received signal amplitude and instantaneous power depend on the phases among RF waves. Since these phases are kept constant during the transmission of each burst, but randomly changed at each one, the received signal amplitude and its instantaneous power at each burst define decorrelated random processes.

Given that the input power is a random process, we statistically analyze its properties.
First, we observe that the expected value of the received power at the sensor node location during the duration of a burst of energy equals to the sum of the contributions ($\mathbb{E} [p_R]  = \sum_i P_{i} /D$), since
\begin{equation}
\mathbb{E} \left[  \cos \left( \phi_i - \phi_j \right) \right] = 0, \; \forall i \neq j.
\end{equation}
The expected value refers to the received power that a sensor node receives on average, if the phases are randomly chosen. In DRAMA, since the phase shifts are randomly varied in time, the temporal average tends to the expected value multiplied by the duty cycle as a consequence of the weak law of large numbers. Then, the input power in temporal average, $P_R$, equals to the sum of the received power from each ET:
\begin{equation}
P_R  = \sum_{i=1}^m P_i .
\end{equation}
We find that 
this guarantees that DRAMA operates as a multiple access scheme, as it does not suffer from interferences in temporal average. Also, we observe that the use duty cycled bursts generates large instantaneous power for the same average power.

The variance of the received power, $\sigma_R$, is given by:
\begin{equation}
\sigma_R^2 = \mathbb{E}[p_R^2] - \mathbb{E}[p_R]^2 = D^{-2} \sum_{\substack{1 \leq i,j \leq m \\ i \neq j}} \mathbb{E} \left[ P_{i} P_{j} \right].
\end{equation}
In order to exemplify this result, if we assume that the received power
from each ET is equal, i.e., $P = P_{1} = P_{2} = \cdots = P_m$, we obtain that the relation between the standard deviation and the average value is given by:
\begin{equation}
\frac{\sigma_R}{P_R} = \sqrt{\frac{m-1}{m}}.
\label{eq:sigma_mu}
\end{equation}
From this example, we observe that as the number of coinciding ETs increases the variation in the input power becomes larger. Therefore, sensors that are located near an ET (and, thus, they mostly receive power from a single source) will receive a less time-varying input power than nodes located at further distances in a real network deployment (See Fig.~\ref{fig:interferers}). It is also interesting to observe that \eqref{eq:sigma_mu} is a monotonically increasing function with maximum value 1. The maximum value is rapidly reached when $m > 6$. As a result, we can infer that after a reasonable number of coinciding ETs (approximately $m >6$) the received power will show similar properties than if we assume that the received power comes from the combination of \emph{infinite} number of power sources. This will be later addressed in the section.

\begin{figure}
  \centering
    \includegraphics[width=0.45\textwidth]{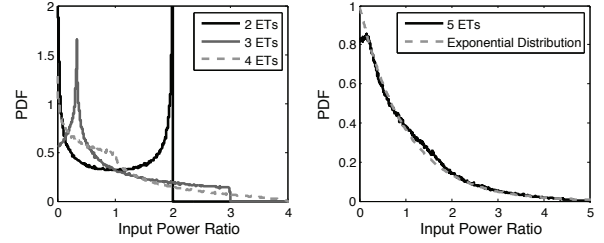}
 \caption{Distribution of the input power ratio for $m =  \{2,3,4,5 \}$ coinciding ETs. The pdf of this ratio tends to an exponential distribution for $m$ sufficiently large.}
 \label{fig:inputpowerpdf}
 \vspace{0.0 cm}
\end{figure}

Unfortunately, as it is found in~\cite{sum_of_sines}, there is no general expression for the pdf, and its calculation results in non-intuitive and very specific formulas for each particular case. In this work, we numerically solve the pdf for different number of simultaneous ETs. Particularly, we show in Fig.~\ref{fig:inputpowerpdf}.a a numerical evaluation of the pdf of the ratio between the received power and its average value when this is generated by 2, 3 and 4 ETs. To derive these pdfs, we have assumed an equal contribution from each ET. As it is observed, the particular case of 2 ETs generates the U-shaped arcsine distribution. Alternatively, it is also shown that when the number of ETs increases, the pdf approaches to the exponential distribution. In Fig.~\ref{fig:inputpowerpdf}.b, the particular case of 5 ETs, with overall $P_R = 1$ is compared to an ideal exponential distribution of $\lambda=1$. As observed, the distribution of the received power is already very similar to the ideal distribution. We consider that, for a number of coinciding RF waves larger than 5, the exponential distribution with $\lambda = P_R$ is a good approximation of the received power at a given burst.

\subsection{Model of the Energy Harvester}
We describe the performance of an energy harvester through its efficiency, here referred as $\eta$. We observe that this efficiency is a function of the input power, as it is described in~\cite{P1110,kaushik-jetcas}, and it is an increasing function of the input power, especially for low input power values. 

We first consider the contribution of the duty cycle. We find that the duty cycle shifts the efficiency curve to lower input values. That is, the efficiently harvested power $P_H$ can be expressed as:
\begin{equation}
P_H = \eta(P_R / D)P_R.
\end{equation}
Notice that dividing the received power by the duty cycle can be expressed as a displacement of the curve of $10 \log(D)$ towards negative values in dB units.

\begin{figure}
  \centering
    \includegraphics[width=0.40\textwidth]{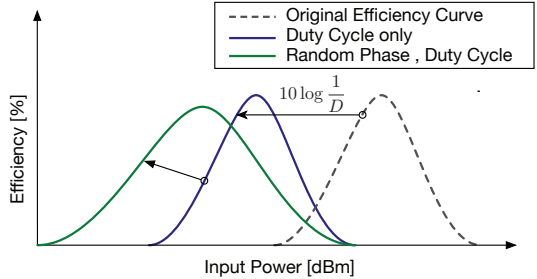}
 \caption{Contribution of the duty cycle and random phase to the eventual efficiency curve. The duty cycle shifts the efficiency curve to lower values of energy by $ -10 \log D$ dB. The random phase equalizes the curve, thus broadening the range of admissible input powers.}
 \label{fig:contributions}
 \vspace{0.0 cm}
\end{figure}

Then we consider the contribution of the random phase by assuming a duty cycle $D=1$, due to the fact that DRAMA intentionally generates both destructive and constructive interferences over time at the sensor node location, the efficiency of the energy harvester also presents a time-varying evolution. As a result, we pursue to find an \emph{average efficiency} value, here referred as $\overline{\eta}$, such that the average harvested power, $P_H$, can be expressed as:
\begin{equation}
P_H = \mathbb{E}[\eta(p_R)p_R] = \overline{\eta}P_R.
\end{equation}
Therefore, to calculate the average efficiency, we find that it can be obtained by:
\begin{equation}
\overline{\eta} = \frac{\int \eta(p_R) f_{P_R} (p_R) \, dp_R}{P_R}
\label{eq:etaavg}
\end{equation}
This integration results in an actual smoothing or equalization of the efficiency curve, which broadens the range of admissible input powers.
To better understand the contributions of the duty cycle and the random phase, we show in Fig.~\ref{fig:contributions} their contributions in tuning and equalizing the efficiency curve.
As it is shown, the duty cycle shifts the efficiency curve down to lower values of input power by $ - 10 \log(D)$ dB, whereas  the random phase broadens the range of admissible input powers.

\section{Numerical Evaluation}
\begin{figure}
  \centering
    \includegraphics[width=0.4\textwidth]{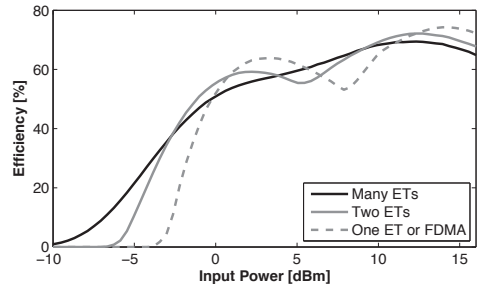}
 \caption{Efficiency curve as a function of the input power for the DRAMA scheme with two and many ETs, compared to the reception of power from a single ET.}
 \label{fig:Powercast}
 \vspace{0.0 cm}
\end{figure}

We first evaluate DRAMA considering the RF-to-DC conversion efficiency of the off-the-shelf energy harvester Powerharvester P1110~\cite{P1110}, from Powercast Corp.
We show the efficiency curve as a function of the input RF power in Fig.~\ref{fig:Powercast} in gray dashed line. It can be observed that at the low input power range the efficiency grows with the input power, whereas it stays approximately constant at the large power range.

In order to show the performance of DRAMA in this scenario, we evaluate \eqref{eq:etaavg} by considering both the exponential distribution (to consider the reception from many ETs) and the arcsine distribution (to consider the reception from two ETs). Provided that the contribution of the duty cycle is reduced to an effective displacement of the curve, the duty cycle of DRAMA has been set to $D=1$ during the numerical evaluation.

The results of these studies are shown in Fig.~\ref{fig:Powercast}. It is shown that the equalizing properties of DRAMA have a remarkable positive impact upon the efficiency at low input powers. In particular, with only two ETs, the energy harvester is able to efficiently operate at lower input powers (from approximately -4~dBm to -6~dBm). In addition, the use of DRAMA not only extends the minimum input power, but it also improves the efficiency. In fact, the efficiency at -3~dBm is increased from approximately 5\% to 40\%. 
However, the major benefits are obtained in the particular case of many ETs. This is mainly due to the fact that the input power is largely randomized with large energetic bursts, which can result in high efficiency in the actual harvested power. In this case, the minimum input power, which enables the conversion of power, is reduced down to approximately -10~dBm.
As a final observation, DRAMA equalizes the efficiency curve, thus providing a consistent efficiency level as a function of the input power. 

\section{Experimental Evaluation}
In this section, we validate the benefit of our approach through an experimental setup consisting of two ETs and one energy harvesting sensor and compare its performance to orthogonal methods for interference cancellation. 

\subsubsection{Experimental Setup}
\begin{figure}
  \centering
    \includegraphics[width=0.45\textwidth]{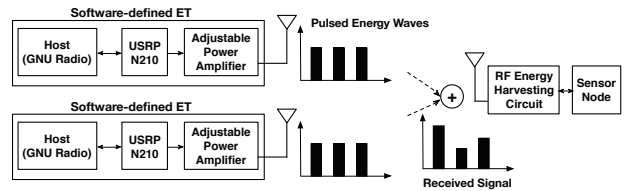}
 \caption{Experimental setup block diagram.}
 \label{fig:blockdiagram}
 \vspace{0.0 cm}
\end{figure}

To build a many-to-one power transmission system, we have employed two USRP Software Radio \cite{ettus} devices and GNURadio open source software to create configurable duty-cycled signals, which emulate the operation of the ETs. 
The USRP devices can be synchronized in time, frequency and phase. 
In this experiment, we have configured the USRPs to generate sine signals at 915~MHz with random phase at each {\em on} duration of the duty cycle, as defined by the DRAMA modulation. The period time for the experiment has been set to $T =42$~ms, with a duty cycle of $D =0.5$.
In addition, we have also configured the USRPs to transmit signals at different frequency bands, centered at 915~MHz, to compare the performance of DRAMA over FDMA, as an example of an orthogonal multiple access method.
The output of USRPs is fed into 3 Watt MPA-0850 RF power amplifiers from RF Bay \cite{rfbay} to amplify USRP signals and generate high power RF waves in a range that can be harvested by P1100 Powercast harvester.
 The block diagram of this experimental setup is shown in Fig.~\ref{fig:blockdiagram}.

At the receiver end, we have connected a 100~mF capacitor at the output of the P1100 Powercast energy harvester. We have calculated the charging time that it takes to charge the capacitor from 0~V to 3.2~V.
The charging time of the capacitor is a valid metric to evaluate the performance of the energy harvester, since it is inversely proportional to the output power of the energy harvester~\cite{kaushik-routing}.

\subsubsection{Experimental Results}
\begin{figure}
  \centering
    \includegraphics[width=0.35\textwidth]{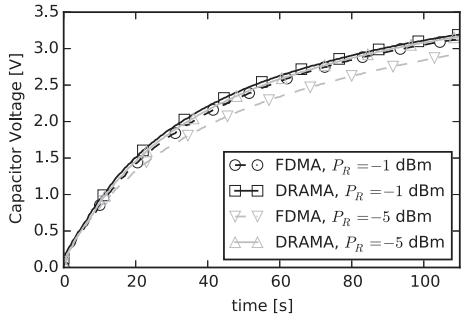}
 \caption{Comparison of the capacitor charging curve between DRAMA and FDMA, for different values of input power.}
 \label{fig:chargefigure}
 \vspace{0.0 cm}
\end{figure}

 Fig.~\ref{fig:chargefigure} shows the charging voltage curve as a function of the time for DRAMA and FDMA for the input RF powers of $P_{R} = \{ -1,\; -5\}$~dBm. 
As it is observed, DRAMA is able to charge the capacitor at a faster speed than FDMA. In addition, the time difference at -5~dBm is very significant. This is consistent with Fig.~\ref{fig:Powercast}, since DRAMA is able to improve the efficiency of the energy harvester efficiency at lower input powers.

\begin{figure}
  \centering
    \includegraphics[width=0.35\textwidth]{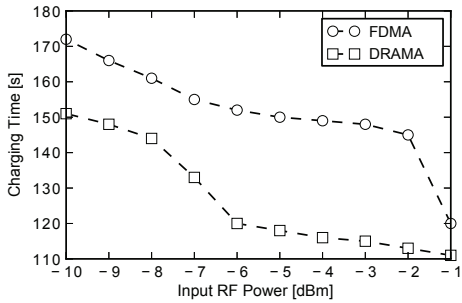}
 \caption{Comparison of the charging times as a function the input power for DRAMA and FDMA.}
 \label{fig:time_results}
 \vspace{0.0 cm}
\end{figure}

In addition to this, Fig.\ref{fig:time_results} shows the charging times as a function of the input power of the two multiple access methods. 
We observe that DRAMA is able to charge the energy storage capacitor at the sensor node at a faster speed. In fact, DRAMA reduces the charging time by 23\%. Alternatively, we observe that the energy harvester requires approximately from 5 to 6 dB less input RF power (i.e., approximately 70\% less input RF power) to charge the capacitor considering equal charging times, when comparing the DRAMA method to FDMA. This reduction in the required input power is very important and results in increasing the power transmission distance. For example, an improvement of 6 dB in the path-loss behavior of a wireless power transmission equals to double the transmission distance in free-space conditions.

\section{Conclusions}
\label{sec:Conclusions}
In this paper, a duty-cycled random-phase multiple access (DRAMA) scheme is proposed. This scheme optimizes the wireless RF power transfer in a many (ETs)-to-many (sensors) transmission, and enables low-power wireless power transmission in a many-to-one configuration. 
The DRAMA is compared to orthogonal multiple access schemes, namely FDMA, through real-testbed experiments.  
We demonstrate that, according to our results, the DRAMA has two-fold benefits: first, it permits the simultaneous action of multiple ETs and second, it improves the efficiency of the power conversion at the node locations. The proposed approach shows 23\% increase in the charging time required to recharge the energy buffer of the sensor.

\section*{Acknowledgements}
This work has been partially funded with the FI-DGR grant from AGAUR and in part by the US National Science Foundation under research grant CNS-1452628.

\balance
\bibliographystyle{IEEEtran}
\bibliography{bibwsn}

\begin{thebibliography}{10}
\providecommand{\url}[1]{#1}
\csname url@samestyle\endcsname
\providecommand{\newblock}{\relax}
\providecommand{\bibinfo}[2]{#2}
\providecommand{\BIBentrySTDinterwordspacing}{\spaceskip=0pt\relax}
\providecommand{\BIBentryALTinterwordstretchfactor}{4}
\providecommand{\BIBentryALTinterwordspacing}{\spaceskip=\fontdimen2\font plus
\BIBentryALTinterwordstretchfactor\fontdimen3\font minus
  \fontdimen4\font\relax}
\providecommand{\BIBforeignlanguage}[2]{{%
\expandafter\ifx\csname l@#1\endcsname\relax
\typeout{** WARNING: IEEEtran.bst: No hyphenation pattern has been}%
\typeout{** loaded for the language `#1'. Using the pattern for}%
\typeout{** the default language instead.}%
\else
\language=\csname l@#1\endcsname
\fi
#2}}
\providecommand{\BIBdecl}{\relax}
\BIBdecl

\bibitem{zane-proceedings}
Z.~Popovic, E.~Falkenstein, D.~Costinett, and R.~Zane, ``Low-power far-field
  wireless powering for wireless sensors,'' \emph{Proceedings of the IEEE},
  vol. 101, no.~6, pp. 1397--1409, June 2013.

\bibitem{naderi-RFMAC}
M.~Naderi, P.~Nintanavongsa, and K.~Chowdhury, ``Rf-mac: A medium access
  control protocol for re-chargeable sensor networks powered by wireless energy
  harvesting,'' \emph{Wireless Communications, IEEE Transactions on}, vol.~13,
  no.~7, pp. 3926--3937, July 2014.

\bibitem{survey-sources}
S.~Sudevalayam and P.~Kulkarni, ``Energy harvesting sensor nodes: Survey and
  implications,'' \emph{IEEE Communications Surveys Tutorials}, vol.~13, no.~3,
  pp. 443 --461, 2011.

\bibitem{powercast}
\BIBentryALTinterwordspacing
P.~Corporation, ``Tx91501 user's manual.'' [Online]. Available:
  \url{http://www.powercastco.com/PDF/TX91501-manual.pdf}
\BIBentrySTDinterwordspacing

\bibitem{P1110}
\BIBentryALTinterwordspacing
------, ``P1110 - 915 {MHz} {RF} {Powerharvester} {Receiver}.'' [Online].
  Available: \url{http://www.powercastco.com/PDF/P1110-datasheet.pdf}
\BIBentrySTDinterwordspacing

\bibitem{kaushik-jetcas}
P.~Nintanavongsa, U.~Muncuk, D.~Lewis, and K.~Chowdhury, ``Design optimization
  and implementation for {RF} energy harvesting circuits,'' \emph{IEEE Journal
  on Emerging and Selected Topics in Circuits and Systems (JETCAS)}, vol.~2,
  no.~1, pp. 24--33, March 2012.

\bibitem{cttc-wpt}
A.~Boaventura, A.~Collado, A.~Georgiadis, and N.~Borges~Carvalho, ``Spatial
  power combining of multi-sine signals for wireless power transmission
  applications,'' \emph{IEEE Transactions on Microwave Theory and Techniques},
  vol.~62, no.~4, pp. 1022--1030, April 2014.

\bibitem{synch-survey}
B.~Sundararaman, U.~Buy, and A.~D. Kshemkalyani, ``Clock synchronization for
  wireless sensor networks: a survey,'' \emph{Ad Hoc Networks}, vol.~3, p.
  281323, 2005.

\bibitem{GPS-timing}
J.~Mannermaa, K.~Kalliomaki, T.~Mansten, and S.~Turunen, ``Timing performance
  of various gps receivers,'' in \emph{Proc. of the Joint Meeting of the
  European Frequency and Time Forum and the IEEE International Frequency
  Control Symposium}, vol.~1, 1999, pp. 287--290 vol.1.

\bibitem{sum_of_sines}
A.~Abdi, H.~Hashemi, and S.~Nader-Esfahani, ``On the pdf of the sum of random
  vectors,'' \emph{IEEE Transactions on Communications}, vol.~48, no.~1, pp.
  7--12, January 2000.

\bibitem{ettus}
\BIBentryALTinterwordspacing
E.~{Research}, ``{N210} {Universal Software Radio Peripheral} ({USRP}).''
  [Online]. Available: \url{http://ettus.com/}
\BIBentrySTDinterwordspacing

\bibitem{rfbay}
\BIBentryALTinterwordspacing
R.~B. {Inc.}, ``{MPA-0850}, {750-950MHz} 5 {W} {RF} {Power} {Amplifier}.''
  [Online]. Available: \url{http://rfbayinc.com/}
\BIBentrySTDinterwordspacing

\bibitem{kaushik-routing}
R.~Doost, K.~Chowdhury, and M.~Di~Felice, ``Routing and link layer protocol
  design for sensor networks with wireless energy transfer,'' in \emph{Proc. of
  the IEEE GLOBECOM}, December 2010, pp. 1 --5.

\end{thebibliography}

\end{document}